\documentclass[12pt]{article}
\usepackage{amssymb}
\newcommand{\be}{\begin{eqnarray}}
\newcommand{\ee}{\end{eqnarray}}
\newcommand{\pa}{\partial}
\renewcommand{\d}{{\rm d}}
\newcommand{\D}{{\rm D}}
\title{\bf Bicomplex formulation and Moyal deformation of  
  $(2+1)$-dimensional \\ Fordy-Kulish systems}

\date{  }

\author{A. Dimakis$^1$ and F. M\"uller-Hoissen$^2$}

\begin{document}

\renewcommand{\theequation} {\arabic{section}.\arabic{equation}}

\maketitle

\begin{center}
 $^1$ Department of Mathematics, University of the Aegean,
        GR-83200 Karlovasi, Greece, dimakis@aegean.gr
\vskip.1cm
 $^2$ Max-Planck-Institut f\"ur Str\"omungsforschung,
        Bunsenstrasse 10, D-37073 G\"ottingen, 
        fmuelle@gwdg.de
\end{center}

\begin{abstract}
Using bicomplex formalism we construct generalizations of Fordy-Kulish 
systems of matrix nonlinear Schr\"odinger equations on two-dimensional space-time 
in two respects. Firstly, we obtain corresponding equations in three space-time dimensions. 
Secondly, a Moyal deformation is applied to the space-time coordinates and the 
ordinary product of functions replaced by the Moyal product in a suitable way. 
Both generalizations preserve the existence of an infinite set of conservation laws. 
\end{abstract}

\section{Introduction}
\setcounter{equation}{0}
A {\em bicomplex} is an ${\mathbb{N}}_0$-graded linear 
space (over ${\mathbb{R}}$ or ${\mathbb{C}}$) $M = \bigoplus_{s \geq 0} M^s$ 
together with two linear maps $\d , \delta \, : \, M^s \rightarrow M^{s+1}$ 
satisfying
\be
   \d^2 = 0 \, , \qquad  \delta^2 = 0 \, , \qquad  
   \d \, \delta + \delta \, \d = 0 \; .   \label{bicomplex_cond}
\ee 
Associated with a bicomplex is the {\em linear equation} 
\be
    \delta \chi = \lambda \, \d \, \chi  
               \label{bc-linear}
\ee
where $\chi \in M^0$ and $\lambda$ is a parameter \cite{DMH00}. 
If it admits a (non-trivial) solution as a (formal) power series 
$\chi = \sum_{r \geq 0} \lambda^r \chi^{(r)}$ in $\lambda$, 
the linear equation leads to 
\be
   \delta \chi^{(0)} = 0 \, , \quad
   \delta \chi^{(r)} = \d \chi^{(r-1)} \, , \quad r=1, \ldots, \infty \; .
\ee 
As a consequence, $J^{(r+1)} = \d \chi^{(r)}$, $r=0, \ldots, \infty$, are 
$\delta$-exact. These elements of $M^1$ may be regarded as generalized conserved 
currents \cite{DMH00}.
\vskip.1cm

In section 2 we start with a trivial bicomplex. A certain ``dressing" (in the 
sense of \cite{DMH00}) then leads to a bicomplex formulation of the Fordy-Kulish 
systems \cite{Ford+Kuli83} of matrix nonlinear Schr\"odinger equations 
(matrix-NLS).\footnote{See also \cite{nls1+1gen}. For some other generalizations 
of the NLS equation see \cite{nls1+1oth}, in particular.} 
More precisely, our approach leads to an extension 
of the latter systems from two to three space-time coordinates $t,x,y$ from 
which the matrix-NLS equations are obtained via the reduction $y=x$.\footnote{A 
different extension of the Fordy-Kulish systems to $2+1$ dimensions obtained 
by replacing the spectral parameter in the $(1+1)$-dimensional systems by 
a new partial derivative appeared in \cite{Atho+Ford87}. 
See also \cite{2+1magnet-gen} for some related work.} 
Our $(2+1)$-dimensional systems turn out to be matrix generalizations of a 
system studied in \cite{2+1magnet,M-NLS} (see also the references given there).
\vskip.1cm
 
 Furthermore, in section 3 deformation quantization \cite{dq} is applied to the 
space-time coordinates. The ordinary commutative product in the algebra $\cal A$  
of smooth functions on $\mathbb{R}^3$ is replaced with the $\ast$-product which 
is defined by 
\be
     f \ast h = {\bf m} \circ e^{i \, {\cal P}/2} (f \otimes h)  
\ee
where ${\bf m} (f \otimes h) = f \, h$ for all $f,h \in {\cal A}$, 
and ${\cal P} : {\cal A} \otimes {\cal A} \rightarrow {\cal A} \otimes {\cal A}$ 
is given by
\be
   {\cal P} = \vartheta_1 \, ( \pa_t \otimes \pa_x - \pa_x \otimes \pa_t )
      + \vartheta_2 \, ( \pa_t \otimes \pa_y - \pa_y \otimes \pa_t )
      + \vartheta_3 \, ( \pa_x \otimes \pa_y - \pa_y \otimes \pa_x )  
           \label{P-example}
\ee
with real deformation parameters $\vartheta_j$, $j=1,2,3$.\footnote{Only one 
of these parameters is actually independent since $\cal P$ is antisymmetric and 
thus degenerate in three dimensions.} 
Under complex conjugation, we have $\overline{f \ast h} = \overline{h} \ast 
\overline{f}$ for functions $f,h$. 
The partial derivatives $\pa_t, \pa_x, \pa_y$ are derivations of the $\ast$-product.
Space-time deformation quantization has been applied recently to various integrable 
models in \cite{Taka00,DMH00nls,DMH00nc}, for example. 
\vskip.1cm

Section 4 deals with the conservation laws of the extended and deformed Fordy-Kulish 
systems. Section 5 gives corresponding generalized ferromagnet equations for the latter 
systems. Section 6 contains some conclusions.

\section{Bicomplex formulation of extended Fordy-Kulish systems}
\setcounter{equation}{0}
We choose the bicomplex space as $M = M^0 \otimes \Lambda$ where $M^0 = C^\infty({\mathbb{R}^3},{\mathbb{C}}^N)$ denotes the set of smooth 
maps $\phi : {\mathbb{R}}^3 \rightarrow {\mathbb{C}}^N$ and 
$\Lambda = \mathbb{C} \oplus \Lambda^1 \oplus \Lambda^2$ is the exterior 
algebra of a $2$-dimensional complex vector space with basis $\tau,\xi$ 
of $\Lambda^1$ (so that $\tau^2 = \xi^2 = \tau \, \xi + \xi \, \tau = 0$). 
$M$ becomes a bicomplex with the maps $\d$ and $\delta$ defined by
\be
   \d \phi &=& \phi_t \, \tau + \phi_x \, \xi  \, ,   \\
   \delta \phi &=& \phi_y \, \tau + (A - a \, I) \, \phi \, \xi  \label{delta}
\ee
where an index denotes a partial derivative with respect to one of the 
coordinates $t,x,y$ on $\mathbb{R}^3$, e.g., $\phi_t = \pa_t \phi$. 
$A$ is a constant $N \times N$ matrix, $I$ the identity matrix, and 
$a \in \mathbb{C}$.\footnote{More precisely, $\d^2$ vanishes identically, 
$\delta^2 =0$ requires $(A-a \, I)_y=0$, and $ \d \, \delta + \delta \, \d = 0$ 
is satisfied iff $(A-a \, I)_t=0$. This still allows an $x$-dependence of 
$A-a \, I$. In the following, we will be interested in the possibility of 
a reduction of the system to two space-time dimensions by setting $y=x$. 
Then $A-a \, I$ has to be constant.} 
By linearity and $\d (\phi \, \tau + \varphi \, \xi) = (\d \phi) \, \tau 
+ (\d \varphi) \, \xi$ (and correspondingly for $\delta$) the maps $\d$ and 
$\delta$ extend to the whole of $M$. Now we apply a ``dressing" to $\d$ as follows,
\be
       \D \phi 
   &=& \d \phi + \delta (L \, \phi) - L \, \delta \phi    \nonumber \\ 
   &=& (\phi_t + L_y \phi) \, \tau + (\phi_x + [A,L] \, \phi) \, \xi
\ee
with an $N \times N$ matrix $L$ and $[A,L] = A \, L - L \, A$.
Besides $\delta^2=0$, also $\delta \D + \D \delta =0$ is identically satisfied. 
The only nontrivial new bicomplex equation is $\D^2 = 0$ which takes the form
\be
    L_{yx} - [ A , L_t ] - [ L_y , [A,L] ] = 0 \; .   \label{D2=0}
\ee
\vskip.1cm

Let us assume that $A$ and $L$ take values in a representation of the Lie algebra 
$\mathfrak{g}$ of a simple Lie group $G$. Let $K$ be a subgroup of $G$ with Lie algebra 
$\mathfrak{k}$, and $\mathfrak{m}$ the vector space complement of $\mathfrak{k}$ in 
$\mathfrak{g}$, so that $\mathfrak{g} = \mathfrak{k} \oplus \mathfrak{m}$ and 
$[\mathfrak{k} , \mathfrak{k} ] \subset \mathfrak{k}$. We assume that the homogeneous 
space $G/K$ is reductive and moreover symmetric, i.e., 
\be
   [ \mathfrak{k} , \mathfrak{m} ] \subset \mathfrak{m}  \, , \qquad  
   [ \mathfrak{m} , \mathfrak{m} ] \subset \mathfrak{k} \; .
\ee 
 For a Hermitian symmetric space with a complex structure $J : \mathfrak{m} 
\rightarrow \mathfrak{m}$, $J^2 = -1$, the following conditions hold (cf \cite{Ford+Kuli83}). 
There is an element $A \in \mathfrak{g}$ such that 
$\mathfrak{k} = \mbox{ker} \, \mbox{ad} A$. For a particular scaling of $A$, we have 
$J = \mbox{ad} A$ and there is a subset $\theta^+$ of the positive root system 
such that $\mathfrak{m} = \mbox{span} \{ e_{\pm \alpha} | \alpha \in \theta^+ \}$ and 
$[A , e_{\pm \alpha}] = \pm i \, e_{\pm \alpha}$ for $\alpha \in \theta^+$. 
Here $e_\alpha$ belongs to the Cartan-Weyl basis. Moreover, $[e_\alpha , e_\beta]
= 0 = [e_{-\alpha} , e_{-\beta}]$ for all $\alpha, \beta \in \theta^+$.
\vskip.1cm

Now we choose $A$ in (\ref{delta}) with the above properties. $A$ is then 
$\mathfrak{k}$-valued. With the decomposition
\be
    L = Q + P
\ee
where $Q \in \mathfrak{m}$ and $P \in \mathfrak{k}$, the $\mathfrak{k}$- and 
$\mathfrak{m}$-part of (\ref{D2=0}) reads, 
respectively,
\be
    P_{xy} &=& [ Q_y , [A,Q]]  \, ,        \label{k-part} \\
    Q_{xy} &=& [ A , Q ]_t + [ P_y , [A,Q] ]
                \label{m-part}  \; .
\ee
According to the above assumptions, $Q$ has a 
decomposition
\be
     Q = Q^+ + Q^-  \, , \qquad    [A , Q^\pm] = \pm i \, Q^\pm  \; .
\ee
Now one finds that (\ref{k-part}) can be integrated with respect to $y$. 
This yields
\be
      P_x = - i \, [ Q^+ , Q^- ]     \label{P-eq}
\ee
(up to addition of a $\mathfrak{k}$-valued matrix which only depends on 
$t$ and $x$ and which we disregard in the following). 
(\ref{m-part}) now leads to 
\be
   i \, Q^\pm_t \mp Q^\pm_{xy} + i \, [P_y , Q^\pm ] = 0 
                \; .  \label{Q-eqs}
\ee
The system of equations (\ref{P-eq}), (\ref{Q-eqs}) constitutes an extension 
of the Fordy-Kulish systems \cite{Ford+Kuli83} to which it reduces when $y=x$. 
In the latter case, (\ref{P-eq}) determines $P_x$ which can then be eliminated 
from (\ref{Q-eqs}). We are then left with the following equations,
\be
   i \, Q^\pm_t \mp Q^\pm_{xx} + [ [Q^+ , Q^-] , Q^\pm ] = 0 
                \; .  \label{Q-nls}
\ee
\vskip.1cm
\noindent
{\em Example.} Let $G = SU(2)$ and $y=x$. The subalgebra of $su(2)$ 
spanned by $A = (i/2) \, \sigma_3$ with the Pauli matrix $\sigma_3$ 
is clearly annihilated by $\mbox{ad} A$ and generates a $U(1)$ subgroup. Let
\be
    Q^+ = \left( \begin{array}{cc} 0 & \psi \\
                                   0 & 0  
                 \end{array} \right) \, , \qquad 
    Q^- = \left( \begin{array}{cc} 0 & 0 \\
                   - \overline{\psi} & 0  
                 \end{array} \right)   
\ee
where $\psi$ is a complex function with complex conjugate $\overline{\psi}$. 
Then the two equations (\ref{Q-nls}) both reduce to the nonlinear Schr\"odinger 
equation
\be
    i \, \psi_t = - \psi_{xx} - 2 \, |\psi|^2 \psi  \; .
\ee
             {  }   \hfill   \rule{5pt}{5pt}
\vskip.1cm

This example is easily generalized \cite{Ford+Kuli83}. Let us consider the Hermitian 
symmetric space $SU(N)/S( U(n) \times U(N-n))$ and choose
\be
    A =  \left( \begin{array}{cc} c_1 \, I_n & 0 \\
                                         0 & c_2 \, I_{N-n}  
                               \end{array} \right)    \label{A-mNLS}
\ee
where $I_n$ is the $n \times n$ unit matrix and $c_1,c_2 \in \mathbb{C}$. With 
\be
    Q^+ = \left( \begin{array}{cc} 0 & q \\
                                   0 & 0  
                 \end{array} \right) \, , \qquad 
    Q^- = \left( \begin{array}{cc} 0 & 0 \\
                          -q^\dagger & 0  
                 \end{array} \right)    \label{Qpm}
\ee
where $q$ is an $n \times (N-n)$ matrix with Hermitian conjugate $q^\dagger$, 
we get
\be
     [ A , Q^\pm ] = \pm (c_1 - c_2) \, Q^\pm  \; .
\ee
The constants $c_1, c_2$ are thus related by $c_1 - c_2 = i$. Since $A$ must be 
traceless, we also have $n c_1 + (N-n) c_2 =0$. Hence
\be
    c_1 = {N-n \over N} \, i \, , \qquad   c_2 = - {n \over N} \, i \; . \label{c1c2}
\ee
The matrix $P$ must have the form
\be
     P = \left( \begin{array}{cc} p & 0 \\
                                  0 & r  
                 \end{array} \right)        \label{Ppr}
\ee
with an $n \times n$ matrix $p$ and an $(N-n) \times (N-n)$ matrix $r$. 
 From (\ref{P-eq})  we obtain the equations
\be
    p_x = i \, q \, q^\dagger \, , \qquad r_x = -i \, q^\dagger \, q   
             \label{p,r-eqs}        
\ee
which are compatible with the unitarity constraints $p^\dagger = - p$ and 
$r^\dagger = - r$, and with $\mbox{tr}(p) + \mbox{tr}(r) = 0$. 
(\ref{Q-eqs}) becomes
\be
    i \, q_t - q_{xy} + i \, ( p_y \, q - q \, r_y ) = 0   \label{q-eq}
\ee
and its Hermitian conjugate. The reduction $y=x$ leads to the matrix 
nonlinear Schr\"odinger equation \cite{Ford+Kuli83}
\be
    i \, q_t - q_{xx} - 2 \, q \, q^\dagger \, q = 0  \; .
\ee
The more general systems determined by (\ref{p,r-eqs}) and (\ref{q-eq}) on 
three-dimensional space-time will be called ``extended matrix-NLS equations" 
in the following.
\vskip.1cm

Other examples of Hermitian symmetric spaces lead to further matrix 
nonlinear Schr\"odinger equations \cite{Ford+Kuli83} and extensions in the 
above sense.

\section{Space-time deformation quantization of the extended matrix-NLS equations}
\setcounter{equation}{0}
In this section we apply a deformation quantization to the algebra of (smooth) functions 
on space-time. The bicomplex $(M,\D,\delta)$ introduced in the previous section then 
generalizes to the deformed noncommutative algebra with the following definition,
\be
       \D \phi 
   &=& \d \phi + \delta (L \ast \phi) - L \ast \delta \phi   \nonumber \\ 
   &=& (\phi_t + L_y \ast \phi) \, \tau + (\phi_x + [A,L]_\ast \ast \phi) \, \xi
\ee
with $[A,L]_\ast = A \ast L - L \ast A$. The only nontrivial bicomplex equation is 
still $\D^2 = 0$ which now takes the form
\be
    L_{yx} - [ A , L_t ]_\ast - [ L_y , [A,L]_\ast ]_\ast = 0 \; .  
\ee
Of course, the $\ast$-commutator does not preserve a Lie algebra structure, in general. 
As a consequence, a decomposition of the last equation like that worked out 
for Hermitian symmetric spaces in \cite{Ford+Kuli83} and the previous section does not 
work, in general. However, in the case of extended matrix-NLS equations only a certain block 
structure of the matrices entering the bicomplex maps is important. Let $\mathfrak{m}^\pm$ 
be the set of all $N \times N$ matrices of the form of $Q^\pm$ in (\ref{Qpm}). Let 
$\mathfrak{k}$ be the set of all block diagonal matrices (like $P$ in (\ref{Ppr})). 
Then we have $\mathfrak{k} \ast \mathfrak{k} \subset \mathfrak{k}$, $\mathfrak{k} 
\ast \mathfrak{m}^\pm \subset \mathfrak{m}^\pm$ and $\mathfrak{m}^\pm \ast \mathfrak{k} 
\subset \mathfrak{m}^\pm$. 
Moreover, since $A$ given in (\ref{A-mNLS}) with (\ref{c1c2}) is constant, we still 
have $[A,P]_\ast = 0$ and $[A , Q^\pm]_\ast = \pm i \, Q^\pm$. Hence, we can proceed 
with the decomposition $L = Q^+ + Q^- + P$ as in the previous section.
The above deformed bicomplex equation now results in the following system,
\be
    p_x = i \, q \ast q^\dagger \, , \qquad r_x = -i \, q^\dagger \ast q
                \label{px-rx}
\ee
and 
\be
    i \, q_t - q_{xy} + i \, ( p_y \ast q - q \ast r_y ) = 0  \; .
\ee
This is a noncommutative version of the corresponding extended matrix-NLS system. 
The equations (\ref{px-rx}) are consistent with unitarity constraints on $p$ 
and $r$, but not with $\mbox{tr}(p) + \mbox{tr}(r) = 0$. In contrast to the 
classical case, these matrices can no longer be taken as Lie algebra valued. 
They have values in the corresponding enveloping algebra instead. 
\vskip.1cm

The reduction $y=x$ of the above system leads to the noncommutative matrix 
nonlinear Schr\"odinger equation
\be
    i \, q_t - q_{xx} - 2 \, q \ast q^\dagger \ast q = 0 
\ee
which is the matrix version of the noncommutative nonlinear Schr\"odinger
equation treated in \cite{DMH00nls}.

\section{Conservation laws for the three-dimensional extensions and deformations 
         of matrix-NLS equations}
\setcounter{equation}{0}
The linear equation associated with the bicomplex underlying the deformed 
extended matrix-NLS equations of the previous section is 
\be
     \delta \chi = \lambda \, \D \chi
\ee
with a parameter $\lambda$. $\chi$ is taken to be an $N \times n$ matrix of 
functions. The linear equation is equivalent to the two equations
\be
  \chi_y &=& \lambda \, ( \chi_t + L_y \ast \chi )  \, ,  \label{linsys1} \\
  (A - a) \, \chi &=& \lambda \, ( \chi_x + [A,L]_\ast \ast \chi )
   \; .    \label{linsys2}
\ee
Let us decompose $\chi$ into an $n \times n$ matrix $\alpha$ and 
an $(N-n) \times n$ matrix $\beta$, 
\be
   \chi = \left( \begin{array}{cc} \alpha \\ \beta \end{array} \right) \; . 
\ee
In order to have a nontrivial solution of $\delta \chi^{(0)} = 0$, we choose 
$a$ as an eigenvalue of $A$. To be more concrete, 
\be
          a = {N-n \over N} \, i     \, , \quad   
 A - a \, I = - i \, \left( \begin{array}{cc} 0 & 0 \\
                                              0 & I_{N-n}  
                            \end{array} \right)  \, , \quad
 \chi^{(0)} = \left( \begin{array}{cc} I_n \\ 0 \end{array} \right) \; .
\ee
 From (\ref{linsys2}) we get
\be
    \alpha_x + i \, q \ast \beta = 0 \, , \qquad
    \beta = i \, \lambda \, ( \beta_x + i \, q^\dagger \ast \alpha )  \; .
\ee
Assuming that $q$ has a left $\ast$-inverse, this implies
\be
     \beta &=& i \, q^{-1}_\ast \ast \alpha_x  \, , \\
  \alpha_x &=& i \, \lambda \, ( \alpha_{xx} - q_x \ast q^{-1}_\ast \ast \alpha_x
               + q \ast q^\dagger \ast \alpha ) \; .   \label{alpha_x}
\ee
 Furthermore, (\ref{linsys1}) leads to
\be
    \alpha_y = \lambda \, ( \alpha_t + p_y \ast \alpha 
               + i \, q_y \ast q^{-1} \ast \alpha_x )  \; .  \label{alpha_y}
\ee
$\alpha$ has a right $\ast$-inverse at least as a formal power series in $\lambda$, 
since at $0$th order it equals $I_n$. Hence there 
are $n \times n$ matrices $\rho$, $\sigma$ and $\zeta$ such that
\be
    \alpha_x = i \, \lambda \, \rho \ast \alpha \, , \quad
    \alpha_t = i \, \lambda \, \sigma \ast \alpha \, , \quad
    \alpha_y = i \, \lambda \, \zeta \ast \alpha \; .    \label{rho...def}  
\ee
Then (\ref{alpha_x}) and (\ref{alpha_y}) lead, respectively, to
\be
    \rho &=& q \ast q^\dagger + i \, \lambda \, ( \rho_x - q_x \ast q^{-1}_\ast 
               \ast \rho ) - \lambda^2 \, \rho \ast \rho \, ,  \label{rho-eq} \\
    \lambda \, \sigma - \zeta 
         &=& i \, p_y - i \, \lambda \, q_y \ast q^{-1}_\ast \ast \rho  \; . 
                          \label{sigma}
\ee
The integrability conditions $\alpha_{xt} = \alpha_{tx}$ and 
$\alpha_{xy} = \alpha_{yx}$ together with 
(\ref{rho...def}) yield
\be
    \rho_t - \sigma_x + i \, \lambda \, [ \rho , \sigma ]_\ast &=& 0 \, , 
                   \label{rho_t}     \\
    \zeta_x - \rho_y - i \, \lambda \, [ \rho , \zeta ]_\ast &=& 0 \; . 
                   \label{zeta_x}
\ee
Differentiation of (\ref{sigma}) with respect to $x$, using (\ref{zeta_x}), 
$i \, p_x = - q \ast q^\dagger$ (cf (\ref{px-rx})) and (\ref{rho-eq}), 
leads to
\be
    \sigma_x = \left( i \, ( \rho_x - q_x \ast q^{-1}_\ast \ast \rho )
     - \lambda \, \rho \ast \rho \right)_y 
     - i \, \left( q_y \ast q^{-1}_\ast \ast \rho \right)_x 
     + i \, [ \rho , \zeta ]_\ast  \; .
\ee
Inserted in (\ref{rho_t}), this yields
\be
    \rho_t 
     + i \, \left( q_y \ast q^{-1}_\ast \ast \rho \right)_x 
     - \left( i \, ( \rho_x - q_x \ast q^{-1}_\ast \ast \rho )
     - \lambda \, \rho \ast \rho \right)_y      
     + i \, [ \rho , \lambda \, \sigma - \zeta ]_\ast = 0 \; .  
                        \label{rho_t-2}
\ee
In terms of the product 
\be
   f \diamond h 
 = {\bf m} \circ \frac{\sin({\cal P}/2)}{{\cal P}/2} (f \otimes h)
                 \label{diam}
\ee
with $\cal P$ defined in (\ref{P-example}), the $\ast$-commutator of two 
functions can be written as follows,
\be
    {1 \over i} [ f , h ]_\ast 
   &=& 2 \, {\bf m} \circ \sin({\cal P}/2) \, (f \otimes h) 
                              \nonumber \\
   &=& \left( f \diamond ( \vartheta_1 \, h_x + \vartheta_2 \, h_y ) \right)_t
       + \left( f \diamond ( -\vartheta_1 \, h_t + \vartheta_3 \, h_y ) \right)_x
       - \left( f \diamond ( \vartheta_2 \, h_t + \vartheta_3 \, h_x ) \right)_y
              \, . \qquad
\ee 
Taking the trace of (\ref{rho_t-2}), using the last formula and (\ref{sigma}), we 
obtain the conservation law
\be
  0 &=& \mbox{tr} \left( \rho - i \, \rho \diamond [( \vartheta_1 \, \pa_x + \vartheta_2 \, \pa_y) 
       ( p_y - \lambda \, q_y \ast q^{-1}_\ast \ast \rho )] \right)_t   \nonumber \\
    & &  + \mbox{tr} \left( i \, q_y \ast q^{-1}_\ast \ast \rho 
       + i \, \rho \diamond [( \vartheta_1 \, \pa_t - \vartheta_3 \, \pa_y ) 
         ( p_y - \lambda \, q_y \ast q^{-1}_\ast \ast \rho ) ] \right)_x   
                                   \label{conserv-law}  \\
    & & + \mbox{tr} \left( \lambda \, \rho \ast \rho  
        - i \, ( \rho_x - q_x \ast q^{-1}_\ast \ast \rho )
        + i \, \rho \diamond [( \vartheta_2 \, \pa_t + \vartheta_3 \, \pa_x )
       ( p_y - \lambda \, q_y \ast q^{-1}_\ast \ast \rho )] \right)_y                
            \; .     \nonumber
\ee 
Expanding $\rho$ in a formal power series in $\lambda$, i.e.,
\be
    \rho = \sum_{r=0}^\infty \lambda^r \, \rho^{(r)} \, , 
\ee
(\ref{rho-eq}) leads to 
\be
   \rho^{(0)} = q \ast q^\dagger \, , \quad
   \rho^{(1)} = i \, q \ast q^\dagger_x \, , \quad
   \rho^{(2)} = - q \ast q^\dagger \ast q \ast q^\dagger - q \ast q^\dagger_{xx}
\ee
and 
\be
   \rho^{(r)} = i \, ( \rho^{(r-1)}_x - q_x \ast q^{-1}_\ast \ast \rho^{(r-1)} )
                - \sum_{s=0}^{r-2} {r-2 \choose s} \, \rho^{(s)} \ast \rho^{(r-2-s)}  
\ee
for $r \geq 2$. Inserting this in the expression
\be
   w = \sum_{r=0}^\infty \lambda^r \, w^{(r)} 
     = \mbox{tr} \left( \rho - i \, \rho \diamond [( \vartheta_1 \, \pa_x 
       + \vartheta_2 \, \pa_y) ( p_y - \lambda \, q_y \ast q^{-1}_\ast \ast \rho )] 
       \right) 
\ee
which appears in the above conservation law, an infinite set of conserved densities 
is obtained, starting with\footnote{These expressions do not involve the $\ast$-inverse 
of $q$ and also apply to solutions for which $q$ is not $\ast$-invertible.}
\be
  w^{(0)} &=& \mbox{tr} \left( q \ast q^\dagger 
              - i \, (q \ast q^\dagger) \diamond [( \vartheta_1 \, \pa_x 
              + \vartheta_2 \, \pa_y) \, p_y ] \right) \, ,          \\
  w^{(1)} &=& \mbox{tr} \left( i \, q \ast q^\dagger_x
              + (q \ast q^\dagger_x) \diamond [( \vartheta_1 \, \pa_x 
              + \vartheta_2 \, \pa_y) \, p_y ]
                \right.                                \nonumber \\
          & & \left. + i \, (q \ast q^\dagger) \diamond [ ( \vartheta_1 \, \pa_x 
              + \vartheta_2 \, \pa_y ) ( q_y \ast q^\dagger ) ]  \right)  \, , \\
  w^{(2)} &=& \mbox{tr} \left( - q \ast q^\dagger \ast q \ast q^\dagger - q \ast q^\dagger_{xx}
              + i \, (q \ast q^\dagger \ast q \ast q^\dagger + q \ast q^\dagger_{xx}) \diamond 
              [ ( \vartheta_1 \, \pa_x + \vartheta_2 \, \pa_y ) \, p_y ]  \right.  \nonumber \\
          &&  \left. - (q \ast q^\dagger_x) \diamond [( \vartheta_1 \, \pa_x 
              + \vartheta_2 \, \pa_y) ( q_y \ast q^\dagger ) ] 
              - ( q \ast q^\dagger ) \diamond [( \vartheta_1 \, \pa_x 
              + \vartheta_2 \, \pa_y) ( q_y \ast q^\dagger_x ) ] \right) \quad
\ee
which in turn can be expanded in (formal) power series in the deformation parameters. 
 For vanishing deformation parameters, the conserved densities $w^{(r)}$ are polynomials 
in $q, q^\dagger$ and their $x$-derivatives, but no $y$-derivatives. This means that 
the conserved densities of an extended matrix-NLS system are the same (as polynomials in 
the fields and their partial derivatives) as those of the corresponding matrix-NLS system 
(which is obtained from the former by setting $y=x$). This is no longer so after 
deformation.

\section{Generalized ferromagnet equations associated with  
         deformed extended Fordy-Kulish systems}
\setcounter{equation}{0}
A gauge transformation of the (deformed) bicomplex considered in section 3 
is a map $g : \mathbb{R}^3 \rightarrow G$ such that
\be
    \D \phi \mapsto \D' \phi' = g^{-1}_\ast \D ( g \ast \phi) \, , \qquad
    \delta \phi \mapsto \delta' \phi' =  g^{-1}_\ast \delta ( g \ast \phi) 
\ee
for all $\phi \in M$ with $\phi' = g \ast \phi$. Such a map leaves the bicomplex 
equations invariant. Let us choose $g$ such that $\D' = \d$. Then
\be
    g^{-1}_\ast \ast g_t = - g^{-1}_\ast \ast L_y \ast g \, , \qquad
    g^{-1}_\ast \ast g_x = - g^{-1}_\ast \ast [A,L]_\ast \ast g
          \label{ginv_dg}
\ee
and, writing ${\cal D}$ instead of $\delta'$, we get
\be
    {\cal D} \phi = (\phi_y + R \ast \phi ) \, \tau
      + ( S - a \, I) \ast \phi \, \xi
\ee
where we introduced the abbreviations
\be
    R = g^{-1}_\ast \ast g_y \, , \qquad  S = g^{-1}_\ast \ast A \ast g \; .
	    \label{RS-def}
\ee
Now ${\cal D}^2 = 0$ reads
\be
    S_y = [S , R ]_\ast
\ee
and $\d {\cal D} + {\cal D} \d = 0$ becomes
\be
    S_t = R_x  \; .   \label{S_t}
\ee
Decomposing $R$ as follows,
\be
     R = U + W \, ,    \label{R-decomp}
\ee
where $g \ast U \ast g^{-1}_\ast \in \mathfrak{m}$ and 
$g \ast W \ast g^{-1}_\ast \in \mathfrak{k}$, 
we find $[S,R]_\ast = [S,U]_\ast$ and thus $[S,S_y]_\ast = [S, [S,U]_\ast ]_\ast$. 
Using $(\mbox{ad} A)^2 = -I$ which implies $(\mbox{ad} S)^2 = -I$, we get 
$[S,S_y]_\ast = - U$ so that (\ref{S_t}) can be written as follows,
\be
      S_t = - ( [S,S_y]_\ast - W )_x  \; .  \label{S-eq}
\ee
 Furthermore, (\ref{R-decomp}) becomes
\be
    g_\ast^{-1} \ast g_y = - [S , S_y]_\ast + W \; .
\ee
(\ref{RS-def}) leads to $S_x = [S , g_\ast^{-1} \ast g_x ]_\ast$ 
(since $A$ is constant) which implies  
\be
    g_\ast^{-1} \ast g_x = - [S , S_x]_\ast  
\ee
by use of $(\mbox{ad} S)^2 = -I$. 
Together with the identity 
\be
   (g_\ast^{-1} \ast g_x)_y - (g_\ast^{-1} \ast g_y)_x
  = [ g_\ast^{-1} \ast g_x , g_\ast^{-1} \ast g_y ]_\ast
\ee
the last two equations lead to
\be
   W_x = [ [S , S_x]_\ast , W ]_\ast + 2 \, [S_x , S_y]_\ast
         - [ [S , S_x]_\ast , [S , S_y]_\ast ]_\ast  \; .
\ee
The second term on the rhs can be rewritten as follows,
\be
     [ [S , S_x]_\ast , [S , S_y]_\ast ]_\ast 
 &=& - [S , [ [S , S_x]_\ast]_\ast , S_y]_\ast 
     + [S , [ [S , S_x]_\ast , S_y]_\ast ]_\ast  \nonumber \\
 &=& [S_x , S_y]_\ast + [S , [ [S , S_x]_\ast , S_y]_\ast ]_\ast  
\ee
using again $(\mbox{ad} S)^2 = -I$.
(\ref{ginv_dg}) implies $g_\ast^{-1} \ast g_x \in g_\ast^{-1} \ast \mathfrak{m} \ast g$ 
and thus $[S , S_x]_\ast \in g_\ast^{-1} \ast \mathfrak{m} \ast g$. It follows 
that $[ [S , S_x]_\ast , S_y]_\ast \in g_\ast^{-1} \ast \mathfrak{k} \ast g$. 
Since $S$ commutes with all elements of $g_\ast^{-1} \ast \mathfrak{k} \ast g$, 
we have $[S , [ [S , S_x]_\ast , S_y]_\ast ]_\ast = 0$. Hence 
\be
    W_x = [S_x , S_y]_\ast + [ [S , S_x]_\ast , W]_\ast  \; .   \label{W-eq}             
\ee
Together with (\ref{S-eq}) this constitutes a $(2+1)$-dimensional matrix 
generalization of the Heisenberg ferromagnet equation, as we shall explain 
below.
%Inserting this expression in (\ref{S-eq}) we get
%\be
%    S_t = - [S , S_{xy}]_\ast + [ [S , S_x]_\ast , W]_\ast \; . 
%\ee

\vskip.2cm
\noindent
{\em Example.} For $N=2$ and $n=1$ we have $S^2 = - {1\over 4} \, I$ and 
$W = u \ast S$ with a function $u$. Using 
\be
 [ [S , S_x]_\ast , u \ast S ]_\ast = [ [S , S_x]_\ast , u ]_\ast \ast S 
  + u \ast [ [S , S_x]_\ast , S ]_\ast 
 = [ [S , S_x]_\ast , u ]_\ast \ast S + u \ast S_x \, ,
\ee
the system (\ref{S-eq}),(\ref{W-eq}) becomes
\be
   S_t = - ( [S , S_y]_\ast - u \ast S )_x \, , \qquad
   u_x = - 4 \, [S_x , S_y]_\ast \ast S + [ [S , S_x]_\ast , u ]_\ast \; .
\ee
In the undeformed (commutative) case, the term $[[S , S_x]_\ast , u]_\ast$ 
disappears. Then we recover a system of equations which has been discussed 
in \cite{2+1magnet} (see also the references given there). 
Its nonlinear Schr\"odinger-type counterpart has been considered in \cite{M-NLS}.
Our equations (\ref{S-eq}),(\ref{W-eq}) thus constitute 
a matrix generalization of this system. Let us choose
\be
   g = \exp \left(- {i \over 2} \sigma_2 \, \varphi \right)
       \exp \left( {i \over 2} \sigma_3 \, t \right) 
     = \left( \begin{array}{cc}
       e^{it/2} \, \cos (\varphi/2) & -e^{-it/2} \, \sin (\varphi/2) \\
       e^{it/2} \, \sin (\varphi/2) &  e^{-it/2} \, \cos (\varphi/2)
       \end{array} \right)
\ee
with a function $\varphi(x,y)$ and the Pauli matrices 
$\sigma_2$ and $\sigma_3$, we get
\be
   S = {i \over 2} \left(\begin{array}{cc} 
       \cos \varphi & - e^{-it} \, \sin \varphi  \\
       - e^{it} \, \sin \varphi & -\cos \varphi
                         \end{array}\right) \; .
\ee
Then $[S_x,S_y] = 0$ and we can choose $W=0$. Now (\ref{S-eq}) 
reduces to the sine-Gordon equation
\be
         \varphi_{xy} = \sin \varphi   \;.
\ee
            {  }   \hfill   \rule{5pt}{5pt}
\vskip.1cm

 For $y=x$, (\ref{ginv_dg}) and the decomposition (\ref{R-decomp}) imply $W=0$. 
Without deformation, (\ref{S-eq}) then reduces to $S_t = - [S,S_{xx}]$ 
which is a matrix generalization \cite{Ford+Kuli83} (see also \cite{Lang+Perl99})
of an equation which describes a one-dimensional continuous spin system 
(Heisenberg ferromagnet) \cite{LRT76}. Its equivalence with the nonlinear 
Schr\"odinger equation was demonstrated in \cite{Zakh+Takh79} (see 
also \cite{Fadd+Takh87}). 
\vskip.1cm

With $S = - \gamma_x$, (\ref{S-eq}) becomes\footnote{An arbitrary matrix $V$ which does 
not depend on $x$ arises as a ``constant of integration". It can be eliminated by a 
redefinition of $\gamma$.} 
\be
     \gamma_t = [ \gamma_x , \gamma_{xy} ]_\ast - W  \; .
\ee
 For $y=x$ we have $W=0$. Without deformation, the last equation is then a 
generalization \cite{Ford+Kuli83} of the {\em Da Rios} equation which describes 
evolution of a thin vortex filament in a three-dimensional fluid \cite{daRi06} 
(see also \cite{Ricca91,Lange99}, in particular). Its equivalence with the nonlinear 
Schr\"odinger equation was first shown in \cite{Hasi72}.

\section{Conclusions}
\setcounter{equation}{0}
Using bicomplex formalism \cite{DMH00}, we obtained an extension of 
the Fordy-Kulish systems of matrix-NLS equations to three space-time dimensions. 
Moreover, corresponding equations on a noncommutative space-time are obtained 
by deformation quantization. We have shown that the resulting equations still 
possess an infinite set of conserved densities. Moreover, there is a 
gauge-equivalent generalized ferromagnet equation for all of these systems.
\vskip.1cm

The Fordy-Kulish systems generalize the nonlinear Schr\"odinger equation, respectively 
the Heisenberg magnet or the Da Rios equation. The latter equations are associated 
with the simplest Hermitian symmetric space $SU(2)/S(U(1)\times U(1))$ in the series 
$SU(N)/S(U(n) \times U(N-n))$. The corresponding {\em extended} Fordy-Kulish system 
associated with this space also reproduces the sine-Gordon equation, as shown in 
the previous section. There may be a way to understand the extended Fordy-Kulish systems 
as generalizations of the sine-Gordon equation in a similar way as they are 
quite obvious generalizations of the NLS equation.  
\vskip.1cm

Certain limits of string, D-brane and M theory generate field theories on noncommutative 
space-times which are obtained by a space-time deformation quantization (see \cite{Seib+Witt99} 
and the references given there). Among the various noncommutative models which arise in these 
and other ways, ``integrable" models will certainly be of special interest because of 
their highly distinguished properties. Although a suitable notion of integrability of a 
noncommutative field theory is not yet at hand, we think that the existence of an 
infinite set of conserved densities should be taken as a partial requirement. In case of 
the deformations considered in this work (see also \cite{DMH00nls,DMH00nc}), this 
requirement is satisfied.

\vskip.2cm
\noindent
{\bf Acknowledgment.} The authors are grateful to Partha Guha for 
discussions and for drawing their attention to Refs. 2 and 13.


\begin{thebibliography}{**}
\bibitem{DMH00} Dimakis A and M\"uller-Hoissen F 2000 
 Bi-differential calculi and integrable models,  
 {\em J. Phys. A: Math. Gen.} {\bf 33} 957--974 \\
 Dimakis A and M\"uller-Hoissen F 2000 Bicomplexes and integrable models 
 {\em J. Phys. A: Math. Gen.} {\bf 33} 6579-6591 
\bibitem{Ford+Kuli83} Fordy A P and Kulish P P 1983 Nonlinear Schr\"odinger equations 
 and simple Lie algebras {\em Commun. Math. Phys.} {\bf 89} 427-443
\bibitem{nls1+1gen} 
 Oh P and Park Q-H 1996 More on generalized Heisenberg ferromagnet models 
 {\em Phys. Lett. B} {\bf 383} 333--338 \\
Terng C L and Uhlenbeck K 1999 Schr\"odinger flows on Grassmannians, 
 math.DG/9901086
\bibitem{nls1+1oth} 
 Fordy A P 1984 Derivative nonlinear Schr\"odinger equations and Hermitian 
 symmetric spaces {\em J. Phys. A: Math. Gen.} {\bf 17} 1235--1245 \\
 Olver P J and Sokolov V V 1998 Non-abelian integrable systems of the nonlinear 
 Schr\"odinger type {\em Inverse Problems} {\bf 14} L5--L8 \\ 
 Tsuchida T and Wadati M 1999 Complete integrability of derivative nonlinear 
 Schr\"odinger equations {\em Inverse Problems} {\bf 15} 1363--1373 \\
 Porsezian K 1997 Completely integrable nonlinear Schr\"odinger type equations
 on moving space curves {\em Phys. Rev. E} {\bf 55} 3785--3788 \\ 
 Porsezian K 1998 Nonlinear Schr\"odinger family on moving space curves: 
 Lax pairs, soliton solution and equivalent spin chains {\em Chaos, Solitons 
 \& Fractals} {\bf 9} 1709--1722 
\bibitem{Atho+Ford87}
 Athorne C and Fordy A 1987 Integrable equations in $(2+1)$ dimensions associated 
 with symmetric and homogeneous spaces {\em J. Math. Phys.} {\bf 28} 2018--2024
\bibitem{2+1magnet-gen} 
 Ishimori Y 1984 Multi-vortex solutions of a two-dimensional nonlinear wave equation 
 {\em Progr. Theor. Phys.} {\bf 72} 33--37 \\
 Cheng Y, Li Y-S and Tang G-X 1990 The gauge equivalence of the Davey-Stewartson 
 equation and $(2+1)$-dimensional continuous Heisenberg ferromagnetic model 
 {\em J. Phys. A: Math. Gen.} {\bf 23} L473--L477 \\
 Konopelchenko B G 1993 {\em Solitons in Multidimensions: Inverse 
 Spectral Transform Method} (Singapore: World Scientific) \\
 Chakravarty S, Kent S L and Newman E T 1995 Some reductions of the self-dual 
 Yang-Mills equations to integrable systems in $2+1$ dimensions 
 {\em J. Math. Phys.} {\bf 36} 763--772  \\
 Radha R and Lakshmanan M 1999 Generalized dromions in the $(2+1)$ dimensional 
 long dispersive wave (2LDW) and scalar nonlinear Schr\"odinger (NLS) equations 
 {\em Chaos, Solitons \& Fractals} {\bf 10} 1821--1824 
\bibitem{2+1magnet} 
 Myrzakulov R, Vijayalakshmi S, Syzdykova R N and Lakshmanan M 1998 
 On the simplest $(2+1)$ dimensional integrable spin systems and their equivalent 
 Schr\"odinger equations {\em J. Math. Phys.} {\bf 39} 2122--2140 \\
 Myrzakulov R, Nugmanova G N and Syzdykova R N 1998 
 Gauge equivalence between $(2+1)$-dimensional continuous Heisenberg ferromagnetic 
 models and nonlinear Schr\"odinger-type equations {\em J. Phys. A: Math. Gen.} 
 {\bf 31} 9535--9545 \\
 Ding Q 1999 The gauge equivalence of the NLS and the 
 Schr\"odinger flow of maps in $2+1$ dimensions {\em J. Phys. A: Math. Gen.} 
 {\bf 32} 5087--5096
\bibitem{M-NLS} 
 Zakharov V E 1980 The inverse scattering method, in {\em Solitons}, 
 eds R K Bullough and P J Caudrey (Berlin: Springer), pp 243--285 \\
 Strachan I A B 1993 Some integrable hierarchies in $(2+1)$ dimensions and their 
 twistor description {\em J. Math. Phys.} {\bf 34} 243--259 \\
 Myrzakulov R, Vijayalakshmi S, Nugmanova G N and Lakshmanan M 1997 
 A $(2+1)$ dimensional integrable spin model: geometrical and gauge equivalent 
 counterpart, solitons and localized coherent structures {\em Phys. Lett. A} 
 {\bf 233} 391--396
\bibitem{dq} Bayen F, Flato M, Fronsdal C, Lichnerowicz A and Sternheimer D 1978 
 Deformation theory and quantization I, II, {\em Ann. Phys.} {\bf 111} 61--151
\bibitem{Taka00} 
 Takasaki K 2000 Anti-self dual Yang-Mills equations on noncommutative spacetime, 
 hep-th/0005194
\bibitem{DMH00nls} Dimakis A and M\"uller-Hoissen F 2000 
 A noncommutative version of the nonlinear Schr\"odinger equation, hep-th/0007015
\bibitem{DMH00nc}
 Dimakis A and M\"uller-Hoissen F 2000  
 Bicomplexes, integrable models, and noncommutative geometry, hep-th/0006005 \\ 
 Dimakis A and M\"uller-Hoissen F 2000 Noncommutative Korteweg-de-Vries equation, 
 hep-th/0007074 \\
 Dimakis A and M\"uller-Hoissen F 2000 Moyal deformation, Seiberg-Witten map, and 
 integrable models, hep-th/0007160, to appear in {\em Lett. Math. Phys.}
\bibitem{Lang+Perl99} 
 Langer J and Perline R 2000 Geometric Realizations of Fordy-Kulish systems, to appear 
 in {\em Pacific J. Math.}
\bibitem{LRT76} Lakshmanan M, Ruijgrok Th W and Thompson C J 1976 On the dynamics of 
 a continuum spin system {\em Physica} {\bf 84A} 577--590
\bibitem{Zakh+Takh79} Zakharov V E and Takhtajan L A 1979 
 Equivalence of the nonlinear Schr\"odinger equation and the equation of a Heisenberg 
 ferromagnet {\em Theor. Math. Phys.} {\bf 38} 17--23
\bibitem{Fadd+Takh87} Faddeev L D and Takhtajan L A 1987 
 {\em Hamiltonian Methods in the Theory of Solitons} 
 (Berlin: Springer)
\bibitem{daRi06} Da Rios L S 1906 Sul moto d'un liquido indefinito con un filetto vorticoso 
 di forma qualunque, {\em Rend. Circ. Mat. Palermo} {\bf 22} 117--135
\bibitem{Ricca91} Ricca R L 1991 Rediscovery of Da Rios equations {\em Nature} {\bf 352} 
 561--562
\bibitem{Lange99} Langer J 1999 Recursion in curve geometry {\em New York Journal of Mathematics}
 {\bf 5} 25--51
\bibitem{Hasi72} Hasimoto H 1972 A soliton on a vortex filament {\em J. Fluid Mech.} 
 {\bf 51} 477--485
\bibitem{Seib+Witt99} Seiberg N and Witten E 1999 String theory
 and noncommutative geometry {\em JHEP} {\bf 9} 32
\end{thebibliography}
\end{document}